\documentstyle[twocolumn,prb,eqsecnum,aps,epsf]{revtex}
\begin{document}
\tighten
\draft

\title{\centerline{Rippled state of double-layer quantum Hall systems}}
\author{C.B. Hanna}
\address{Department of Physics, Boise State University,
Boise, Idaho~~83725}

\date{June 2, 2002}
\maketitle

\begin{abstract}

\vspace{-0.3in}

The incommensurate phase of a bilayer quantum Hall state is found to have
a ``rippled'' dipole charge density whenever the layers are unbalanced.
This tunable dipole-density-wave instability could be detected by
sensitive capacitance measurements and by anisotropic transport.
We demonstrate this explicitly by carrying out a Hartree-Fock calculation
of the layer densities and capacitance for a double-layer quantum Hall state
at a total filling factor of 1.

\end{abstract}


\section{Introduction}
\label{sec:intro}

The combination of reduced dimensionality and strong interparticle
interactions can have spectacular effects on the nature of the
ground-state and dynamical properties of many-particle systems.
This is especially evident in the fractional quantum Hall
regime,\cite{tsui,prange}
where a strong magnetic field is applied perpendicular to a
two-dimensional electron gas at very low temperatures.
The powerful magnetic field quenches the kinetic energy of
the electrons, so that interactions between electrons dominate
the energetics.
The result is a highly correlated, incompressible quantum-liquid
ground state that supports fractionally charged
excitations.\cite{laughlin}
Even at integer filling factors (especially $\nu_{\rm T}$=$1$),
the combination of the quenched kinetic energy of interacting electrons
plus extra electronic degrees of freedom (i.e., spin or multiple layers)
can give rise to correlated ground states that support remarkable
topological excitations, such as charged skyrmions.\cite{sondhi,gmchap}

Another important class of systems in which reduced dimensionality
and interactions strongly affect the ground state and dynamics
are systems with charge-density-wave or spin-density-wave
ground states,\cite{gruner}
and also systems that exhibit commensurate-incommensurate (CI)
transitions.\cite{bak,dennijs}
Systems with charge- or spin-density-wave ground states occur
most famously (although not exclusively) in quasi-one-dimensional materials,
where the reduced dimensionality enhances fluctuation effects
and leads to broken-symmetry ground states.\cite{gruner}
Systems exhibiting CI transitions have broken translational
symmetry states with rich structures, most notably
arrays of domain walls that arise from the competition between
interparticle interactions and external periodic potentials.\cite{bak,dennijs}

Double-layer quantum Hall (2LQH) systems can show both types of behavior:
they support unusual fractionally charged topological
excitations,\cite{yang,moon}
and they apparently exhibit a CI transition to a state
with broken translational symmetry in the presence
of a sufficently strong in-plane magnetic field.\cite{yang}
It is interesting that in 2LQH systems,
broken translational symmetry can coexist with the hidden
off-diagonal long-range order\cite{order} characteristic of
quantum Hall states.
The evidence for a CI transition in 2LQH systems is,
however, indirect.
Activation energy measurements show that a sufficiently large
in-plane magnetic field drives a transition between
two different types of many-body ground states,\cite{murphy}
and the strength of the in-plane magnetic field and the size
of the gaps are consistent with the CI scenario.\cite{yang,murphy}
But in order to establish the nature of the competing quantum Hall
ground states and the transition between them, additional
types of experimental measurements are needed.
Various types of experimental signatures of the CI scenario have
been proposed, including predictions of the form of the energy
gap,\cite{read} a predicted Kosterlitz-Thouless transition,\cite{moon}
anisotropic transport in narrow samples,\cite{mullen}
and the field dependence
of the in-plane magnetization.\cite{ep2ds12,hannasl}
Here we propose that the CI transition in 2LQH systems
could be studied directly by capacitance measurements,
and by anisotropic transport produced by the dipole-density wave
described below.

This paper examines a novel effect of unequal layer densities
on the CI transition in 2LQH systems.
(Interestingly, unequal layer densities can actually enhance the
stability of the 2LQH state at a total filling factor
$\nu_{\rm T}$=$1$.\cite{hannamar97,jogunequal})
It is found that when the layer densities are not equal,
they go from being uniform in the commensurate phase
to becoming ``rippled'' in the incommensurate phase;
this allows the CI transition to be detected
by sensitive capacitance measurements.\cite{hannamar97}
The layer imbalance can be produced in two ways: most commonly
by an external bias, but perhaps also spontaneously in a tilted sample
with an unusually small capacitive charging energy and 
a sufficiently large interlayer tunneling.\cite{radzcant,abolglobal}
We will focus here on bias-driven imbalance, since it is easier
to achieve experimentally.
In addition, capacitive techniques provide a quantitative measure
of the interlayer exchange and pseudospin stiffness in 2LQH systems.
This is illustrated in the following sections by a Hartree-Fock
calculation of the layer densities and capacitances
for a 2LQH state at a total filling factor $\nu_{\rm T}$=$1$.
Similar effects should, in principle, occur at other filling factors,
although $\nu_{\rm T}$=$1$ is probably most promising for
experimental observation.

\section{Rippled state}
\label{sec:m_z}

Interlayer Coulomb interactions at low filling factors can stabilize
2LQH states when the layer spacing
is comparable to the separation between electrons within the
layers.\cite{dlexpt}
Even at a total ``integer'' filling factor $\nu_{\rm T}$=$1$, experiments
indicate that the quantum Hall ground states are stabilized by
Coulomb interactions and do not require interlayer tunneling for their
existence.\cite{dlexpt,ahmpaper}
Further evidence of the rich variety of 2LQH states
at $\nu_{\rm T}$=$1$ comes from measuring the effects of
an in-plane magnetic field, which induces a transition between two
types of quantum Hall ground states.\cite{murphy}
These effects have been discussed in terms of an unusual broken-symmetry
quantum Hall ground states that exhibit spontaneous interlayer (phase)
coherence (SILC).\cite{yang,moon}

At sufficiently small layer separation, a 2LQH system
is an unusual quantum itinerant ferromagnet.\cite{gmchap,yang,moon}
The SILC 2LQH quantum ferromagnet exhibits a rich variety of ground states,
phase transitions, and charged and neutral excitations.\cite{moon,bigyang}
Murphy {\it et al.} investigated the effect of an in-plane
magnetic field $B_\parallel$ on 2LQH systems,
and found evidence of a phase transition between two competing
QH ground states at a critical value
$B_\parallel = B_{\rm c}$.\cite{murphy}
These two ground states have been explained theoretically\cite{yang}
by showing that
application of a sufficiently strong parallel magnetic field
$B_\parallel > B_{\rm c}$ produces a soliton-lattice (SL)
ground state in the incommensurate phase of the 2LQH system.
Recent measurements of the interlayer tunneling conductivity in
bilayer quantum Hall samples have provided dramatic evidence
for interlayer phase coherence.\cite{spielman}

\subsection{Effective Hamiltonian}
\label{subsec:effham}

Formally, it is simplest to obtain the ground-state characteristices of
the SILC 2LQH state from the energy per unit area within a
gradient approximation\cite{yang} in which the pseudospin
${\bf m}({\bf r})$ is assumed to vary slowly on the scale of the
magnetic length $\ell$.
In doing so,
it is convenient to specify the order parameter ${\bf m(r)}$
in terms of two quantities:
$m_z({\bf r})$, the local difference in layer occupancies, and
$\theta({\bf r})$, the projected angle of ${\bf m}$ in the $xy$ plane
measured with respect to the $x$ axis.
For constant $m_z$=$\nu_1$-$\nu_2$ and in-plane magnetic field
${\bf B}_\parallel$, the energy per unit area of the SILC 2LQH state
has the form\cite{yang,moon,bigyang}
\begin{eqnarray}
\label{eq:edensity}
{\cal{E}}
  = \frac{1}{2\pi\ell^2} \lbrack &-&
    t\cos\tilde{\theta} + \frac{1}{2} \rho_{\rm s} 2\pi\ell^2
    \left( \nabla \tilde{\theta} - {\bf Q} \right)^2 \\ \nonumber
&+& \frac{U}{4} m_z^2
 -  \frac{1}{2} V_g m_z
    \rbrack ,
\end{eqnarray}
where ${\cal{E}}$ has been expressed in terms of
$\tilde{\theta} = \theta + {\bf Q}\cdot{\bf r}$,
${\bf Q} \equiv {\bf \hat{z} \times B_\parallel}2\pi d/\phi_0$,
$d$ is the interlayer spacing, and $\phi_0=h/e$ is the
magnetic-flux quantum.
Mean-field equations for $\tilde{\theta}$ and $m_z$ are obtained
by minimizing  Eq.~(\ref{eq:edensity}) with respect to those same
quantities.

The first two terms of Eq.~(\ref{eq:edensity}) constitute the
Pokrovsky-Talapov (PT) model\cite{bak,dennijs,yang,pt} with coefficients
that, in the Hartree-Fock approximation (HFA), depend on $m_z$ according to
\begin{eqnarray}
\label{eq:deftrhos}
t &\equiv& t_0 \sqrt{1-m_z^2} e^{-Q^2\ell^2/4}
= t_0 \sqrt{4\nu_1\nu_2} e^{-Q^2\ell^2/4} \\ \nonumber
\rho_{\rm s} &\equiv& \rho_0 (1-m_z^2)
= 4\nu_1\nu_2\rho_0 ,
\end{eqnarray}
where $t_0$ is the tunneling-matrix element when $\nu_1$=$\nu_2$;
it is equal to half the symmetric-antisymmetric gap $\Delta_{SAS}$.
In the presence of a parallel magnetic field,
\begin{equation}
t_0 \rightarrow t_0 \exp(-Q^2\ell^2/4) ,
\end{equation}
which is a single-body effect.\cite{hu}
The interlayer pseudospin stiffness when the layers are balanced is
\begin{equation}
\rho_0 = \frac{e^2}{4\pi\epsilon\ell} \frac{1}{16\pi}
         \int_0^\infty dx x^2 e^{-x^2/2} e^{-xd/\ell}
\end{equation}
in the HFA.
The value of $\rho_s$ will be reduced due to
quantum fluctuations\cite{moonfluct,yogfluct} and finite-thickness effects.
By adjusting the front and back gate voltages of the sample,
$\nu_1$ and $\nu_2$ may be varied (with $\nu_{\rm T}\equiv\nu_1+\nu_2=1$),
thereby allowing $t$ and $\rho_{\rm s}$ to be adjusted.\cite{hannamar97}

The third term (quadratic in $m_z$) in the energy density
[Eq.~(\ref{eq:edensity})] is a
capacitive charging energy that favors equal layer densities.\cite{yang}
The capacitive energy $U$ is given in terms of the electrostatic
Hartree energy ($\bar{D}_1$) and the intralayer ($\bar{E}_0$) and
interlayer ($\bar{E}_1$) exchange energies by
\begin{eqnarray}
U &=& \bar{D}_1 - \bar{E}_0 + \bar{E}_1 \\ \nonumber
\bar{D}_1 &=& \frac{e^2}{4\pi\epsilon\ell} d/\ell \\ \nonumber
\bar{E}_j &=& \frac{e^2}{4\pi\epsilon\ell} I_j
          \equiv \frac{e^2}{4\pi\epsilon\ell}
                 \int_0^\infty dx e^{-x^2/2} e^{-xdj/\ell} ,
\end{eqnarray}
where the exchange integrals $I_j$ have been evaluated in the HFA.
Most treatments of the SILC 2LQH state have been for equal layer densities
($m_z$=$0$), since there is a significant cost in capacitive charging energy
to unbalance the layers.\cite{yang}
However, an application of back and front gate voltages allows charge to be
transferred from one layer to another, giving rise to a tunable nonzero
value for $m_z$ in the 2LQH ground state.\cite{hannamar97}
The effects of charge-transfer imbalance were studied both
theoretically and experimentally, and it was found that charge imbalance
can actually {\em increase} the stability of the 2LQH
state.\cite{hannamar97,hamilton,sawada}

We shall estimate the numerical values of our results
for a hypothetical ``typical'' GaAs
($m^*\approx 0.07m_e$, $\epsilon_r\approx 13$) 2LQH sample,\cite{hannasl}
with a total density $n_T$=$1.0\times 10^{11}~\rm{cm}^{-2}$,
a layer (midwell to midwell) separation $d$=$20$~nm,
and a tunneling energy $t_0$=0.5 meV ($\Delta_{\rm SAS} = 11.6$ K).
Such a sample would have $\ell\approx 12.6$~nm, $d/\ell\approx 1.6$,
$\hbar\omega_{\rm c}\approx 6.9$~meV for $\nu_{\rm T}$=$1$, and
$e^2/4\pi\epsilon\ell\approx 8.8$~meV.
In the HFA,
$\rho_0\approx 0.03$~meV, $\bar{D}_1 = 14$~meV, and $U = 7.4$~meV.

The effect of the gate voltages is described by the last ($V_g$)
term in the energy density in terms of effective filling factors
$\bar{\nu}_{\rm F}$ and $\bar{\nu}_{\rm B}$
for the front (F) and back (B) gates:
\begin{equation}
\label{eq:vg}
V_g = (\bar{\nu}_{\rm F} - \bar{\nu}_{\rm B}) \bar{D}_1 .
\end{equation}
The effective filling factors $\bar{\nu}_\alpha$ (where $\alpha$=F,B)
are defined by the the electric fields $E_\alpha$ produced by
the front and back gates through Gauss' law
$E_\alpha = e\bar{\nu}_{\rm \alpha}/2\pi\ell^2\epsilon$,
where $\epsilon$ (approximately $13\epsilon_0$ for GaAs)
is the dielectric constant appropriate to the 2LQH sample.

\subsection{Parallel magnetic field}
\label{subsec:parallel}

When $Q\ne 0$, the pseudospin stiffness $\rho_s$ competes with
the rotating Zeeman pseudofield $t\cos\tilde{\theta}$ to determine
the spatial orientations of the pseudospins.
Minimizing ${\cal{E}}$ in Eq.~(\ref{eq:edensity}) with respect to
$\tilde{\theta}$ gives the two-dimensional sine-Gordon equation
\begin{equation}
\xi^2 \nabla^2 \tilde{\theta} = \sin\tilde{\theta} ,
\end{equation}
where the width of the soliton is proportional to the length,
\begin{equation}
\label{eq:xi}
\xi = \sqrt{2\pi\rho_{\rm s}/t}
    = \xi_0 \left(1-m_z^2\right)^{1/4} ,
\end{equation}
which for our hypothetical sample gives $\xi_0\approx 17$~nm.
The soliton width $\xi$ sets the scale for spatial variations of the
pseudospin ${\bf m}({\bf r})$; thus the condition for the validity of
the gradient approximation is that $\xi$ be significantly larger than
the magnetic length $\ell$.

For sufficiently small $Q$, Eq.~(\ref{eq:edensity})
is minimized by $\tilde{\theta}({\bf r})=0$.
This is called the commensurate (C) phase,
and in this phase the pseudospins align themselves with
the rotating Zeeman pseudofield, so that $\theta=-{\bf Q}\cdot{\bf r}$.
However, above a critical value of $B_\parallel$ corresponding to
$Q_c=4/(\pi\xi)=(4/\pi\ell)\sqrt{t/2\pi\rho_s}$,
it becomes energetically favorable to produce
dislocation lines (solitons).
The soliton widths are of order $\xi$.
Solitons proliferate rapidly for $Q>Q_c$ (incommensurate phase)
because they repel each other only very (exponentially) weakly.
The resulting array of solitons breaks the translational symmetry
of the 2LQH ground state by forming a SL.
For large $Q\gg Q_c$, the rapidly varying tunneling phase factor
causes the pseudospins to behave (nearly) as if $t$=$0$.
In the HFA, the critical value of the in-plane field
varies with the layer filling factors like
$Q_c\propto (1-m_z^2)^{-1/4}$; thus tuning the layer filling
factors ($m_z$) via gate voltages allows the location ($Q_c$)
of the CI transtion to be fine tuned.\cite{hannamar97}

The density of soliton lines in the incommensurate phase is proportional
to the soliton wave vector $Q_s=2\pi/L_s$, where $L_s$ is the spacing
between solitons in the SL.
The soliton density (proportional to $Q_s$) is calculated as a function of
the in-plane field (proportional to $Q$) via two equations involving an
intermediate parameter $\eta$.
The parameter $\eta$ is defined by\cite{bak,hannasl}
\begin{equation}
\label{eq:qsqc}
Q_s/Q_c = (\pi/2)^2/[\eta K(\eta)] ,
\end{equation}
and approaches 1 near the CI transition
($Q_s\rightarrow 0$) and goes to 0 deep in the
incommensurate phase ($Q_s\rightarrow\infty$).
Here  $K(\eta)$ is the complete elliptic integral of the first kind.\cite{gr}

\begin{figure}[h]
\epsfxsize3.5in
\centerline{\hspace{0.3in}\epsffile{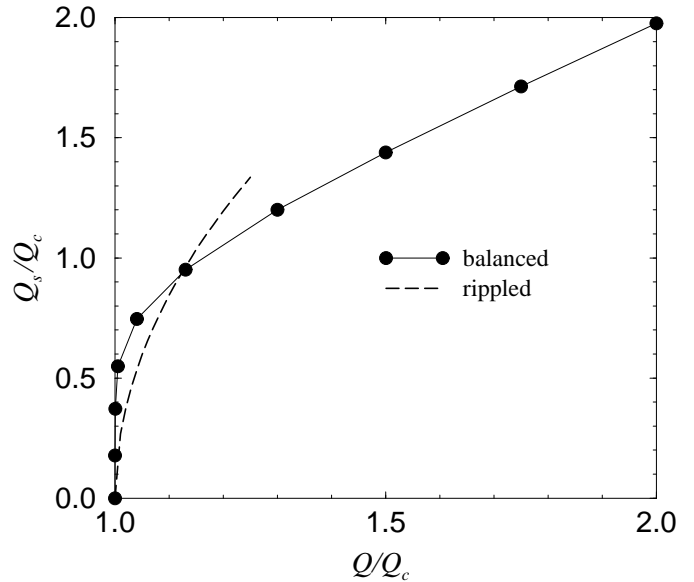}}
\caption{
Soliton density (times $2\pi$) vs in-plane magnetic field
for balanced layers ($m_z$=0, solid curve) and for the rippled state
($m_{z0}$=0.5, dashed curve).
The soliton density rises abruptly in the incommensurate phase,
$Q>Q_{\rm c}$.
There are no solitons in the commensurate phase, $Q<Q_{\rm c}$.}
\label{fig:qs}
\end{figure}

When the layers are balanced ($\nu_1$=$\nu_2$), minimizing
the total energy with respect to $Q_s$ gives\cite{bak,hannasl}
\begin{equation}
\label{eq:qqc}
Q/Q_c = E(\eta)/\eta ,
\end{equation}
where $E(\eta)$ is the complete elliptic integral of
the second kind.\cite{gr}
Equations (\ref{eq:qqc}) and (\ref{eq:qsqc}) together determine
the soliton density (or $Q_s$) as a function of the in-plane magnetic
field (or $Q$).
$Q_s/Q_c$ is plotted as a function of $Q/Q_c$ in Fig.~1,
for the balanced case\cite{hannasl} ($m_z$=0, solid curve)
and for the rippled state ($m_z$=0.5, dashed curve).
Note the abrupt rise in soliton density for $Q>Q_c$.
The compressional stiffness $K_1$ of the SL\cite{hannasl}
is proportional to the slope of the $Q_s$ versus $Q$ curves in
Fig.~\ref{fig:qs} (see Ref. \onlinecite{hannasl}):
\begin{equation}
\label{eq:k1def}
\frac{\rho_s}{K_1} = \frac{\partial Q_s}{\partial Q} .
\end{equation}
As expected, $K_1$ vanishes as $Q \rightarrow Q_c$.

\subsection{Rippled layer imbalance}

\begin{figure}[h]
\epsfxsize3.5in
\centerline{\epsffile{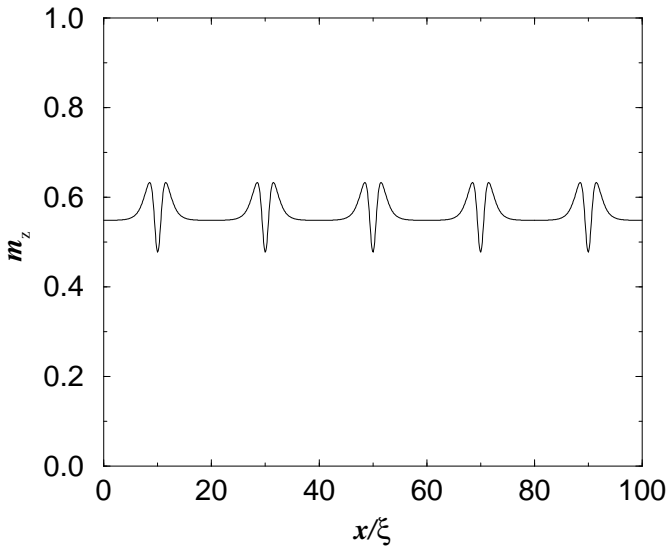}}
\caption{Rippled layer imbalance $m_z(x)$ vs position
for a soliton lattice with spacing $L_s/\xi$=20 between soliton lines.
We have used the parameters for the ``typical'' sample described
in the text, and taken $m_{z0}$=0.5.
For these parameters, $\bar{m}_{z1}$=0.048.}
\label{fig:ripple}
\end{figure}

Minimizing Eq.~(\ref{eq:edensity}) with respect to variations in $m_z$
gives
\begin{equation}
\label{eq:mz}
m_z = \frac{ V_g }
           { U + [2t\cos\tilde{\theta} - 4\pi\ell^2 \rho_{\rm s}
               (\nabla \tilde{\theta} - {\bf Q})^2]/(1-m_z^2) } .
\end{equation}
This equation determines (in the Hartree-Fock gradient approximation)
the filling factor of each layer for the $\nu_{\rm T}=1$ SILC state.
For sufficiently small in-plane magnetic fields
($Q<Q_{\rm c}$), $\tilde{\theta}$=0 (the commensurate state),
and the assumption that $m_z$ is constant is
self-consistent, provided that $t\ll U$.\cite{bigyang}
However, when $Q\ge Q_{\rm c}$,
the quantum Hall ground state breaks translational invariance:
$\tilde{\theta}$ is {\em not} spatially uniform
and the soliton-lattice state is obtained.\cite{yang}
When $Q\ge Q_{\rm c}$ and $\bar{\nu}_1\ne\bar{\nu}_2$,
Eq.~(\ref{eq:mz}) shows that $m_z$ also breaks translation invariance,
and one obtains ``rippled'' layer densities.
Thus uniform $m_z$ is not consistent with the broken translation
symmetry of $\tilde{\theta}$=0 in the incommensurate phase, when
$\bar{\nu}_1\ne\bar{\nu}_2$. 
The resulting behavior of $m_z({\bf r})$ is illustrated in
Fig. \ref{fig:ripple}.

It is important to note that Eq.~(\ref{eq:mz}) is valid
even in the incommensurate state of spatially varying $\tilde{\theta}$,
but only up to first order in $t/U$.
Since $t/U$ is small in the bilayer samples studied so far,
this assumption is not too restrictive.  From now on, we shall work only
to lowest non-trivial order in $t/U$, and expand $m_z$ as described in the
Appendix.  Thus we shall not include the effect of the rippling of
$m_z$ on the soliton width $\xi$, since this would produce only small
corrections (i.e., of higher order in $t/U$) to the results presented here.
The soliton width $\xi$ in Eq.~(\ref{eq:xi}) is therefore computed using
$m_z \approx m_{z0}$, where $m_{z0}$ is the layer imbalance in the absence
of interlayer tunneling, as defined in the Appendix.
The value of $m_{z0}$ depends on the gate voltage $V_g$, but not on
the ``rippling'' effect (which is of higher order in $t/U$).

Because $m_z$=$\nu_1$-$\nu_2$ is associated with differences in layer
electron densities, the rippling has the effect of associating an
electric-dipole density with each soliton.
When the solitons are separated ($L_s>\xi$),
the dipole-moment per unit length is (see the Appendix)
\begin{equation}
\label{eq:elambda}
\frac{\delta p}{\delta y} =
-\frac{ed\xi_0}{2\pi\ell_0^2} \frac{2t_0}{U}
\frac{m_{z0}}{(1-m_{z0}^2)^{1/4}}
(4Q/Q_c - 3) ,
\end{equation}
which has a value of about $-0.10 e$ when $m_{z0}$=0.5 and $Q$=$Q_c$.
Because the solitons have associated dipole moments,
the dominant interactions between solitons when the solitons are
separated will be their dipole-dipole repulsion.
In the limit that the solitons are well separated ($L_s\gg\xi,d$),
the interaction per unit length between two solitons separated
by a distance $x$ is
\begin{equation}
\label{eq:vl}
\frac{{\cal{V}}}{L} = \frac{(\delta p/\delta y)^2}{2\pi\epsilon x^2} ,
\end{equation}
so that the solitons repel each other with a force that falls off
with distance as an inverse power (for $\nu_1\ne\nu_2$), rather than
exponentially (for $\nu_1=\nu_2$).
Summing all the the dipole-dipole soliton interactions gives,
in the thermodynamic limit,
\begin{equation}
\frac{{\cal{V}}}{L_xL_y} =
\frac{\pi}{12} \frac{(\delta p/\delta y)^2}{\epsilon L_s^3} =
\frac{1}{96\pi^2} \frac{(\delta p/\delta y)^2}{\epsilon} Q_s^3 ,
\end{equation}
which is proportional to $(t/U)^2$ and is therefore small in magnitude.
However, near the CI transition, the solitons are well separated and
dipole interactions dominate the repulsions between the solitons,
which has a strong effect on the compressional stiffness $K_1$ of the SL.
The relation between the wave vector $Q$ and the parameter $\eta$
is obtained by minimizing the total energy per unit area with respect
to $Q_s$ at fixed $Q$ (Ref. \onlinecite{hannasl});
when the layers are balanced, Eq.~(\ref{eq:qqc}) results,
and Eqs. (\ref{eq:qsqc}) and (\ref{eq:qqc})
may be combined to obtain $Q_s$ as a function of $Q$.
When the layers are imbalanced,
Eq.~(\ref{eq:qqc}) acquires an additional term due to
the dipole interactions between solitons,
\begin{eqnarray}
\label{eq:qsrip}
Q/Q_c &=& E(\eta)/\eta +
\frac{1}{\rho_s Q_c} \frac{\partial}{\partial Q_s}
\frac{{\cal V}}{L_xL_y}
\\ \nonumber &=&
E(\eta)/\eta + C (Q_s/Q_c)^2 ,
\end{eqnarray}
where $C \sim 0.14$ for the hypothetical ``typical'' sample with
$m_{z0}$=0.5 and $Q$=$Q_c$.
Near the CI transition Eq.~(\ref{eq:qsrip}) gives
\begin{eqnarray}
\label{eq:newqs}
Q_s/Q_c &\approx& \sqrt{(Q/Q_c-1)/C} ,
\\ \nonumber
K_1/\rho_s &\approx& 2 \sqrt{C(Q/Q_c-1)} .
\end{eqnarray}
The corresponding formula for the balanced ($\nu_1$=$\nu_2$) case
are quite different: $Q_s \sim -1/\ln(Q-Q_c)$ and $K_1 \sim (Q-Q_c)$,
up to logarithmic corrections.
Nonetheless, $K_1$ (and, by definition, $Q_s$) vanishes at the
CI transition which has important consequences for the capacitance
near the CI transition.

\subsection{Topological charge}

We also note that when the layers are imbalanced, the soliton
lines acquire a charge density when they are tilted or sheared.
In the lowest Landau level, the pseudospin textures with
topological charge possess an electric charge.  The electric-charge
density $\delta\rho({\bf r})$ is just the topological-charge density
(i.e., the Pontryagian index density\cite{moon}) times $-e\nu_T$,
\begin{equation}
\label{eq:pontry}
\delta\rho = \frac{e\nu_T}{8\pi} \epsilon_{ij}
{\bf m} \cdot (\partial_i {\bf m} \times \partial_j {\bf m}) ,
\end{equation}
where $\epsilon_{ij}$ is the totally antisymmetric tensor of rank 2,
and we sum over $i$ and $j$ over the values 1 and 2 ($x$ and $y$).
The total filling factor is defined as $\nu_T$=$\nu_1$+$\nu_2$.
Expression Eq.~(\ref{eq:pontry}) in terms of $m_z$ and $\theta$ using
\begin{equation}
m_x = \sqrt{1-m_z^2} \cos\theta ,
\qquad
m_y = \sqrt{1-m_z^2} \sin\theta
\end{equation}
gives
\begin{eqnarray}
\delta\rho({\bf r}) &=& \frac{e\nu_T}{4\pi}
(\partial_x \theta \partial_y m_z - \partial_x m_z \partial_y \theta)
\\ \nonumber &=& \frac{e\nu_T}{4\pi}
[(\partial_x \tilde{\theta} -Q) \partial_y m_z
- \partial_x m_z \partial_y \tilde{\theta}] ,
\end{eqnarray}
where we have taken ${\bf Q}$ to be along the ${\bf \hat{x}}$ direction.
If we rotate the soliton lines by taking
\begin{equation}
\tilde{\theta}_0(x) \rightarrow \tilde{\theta}_0(\alpha x + \beta y) ,
\end{equation}
and also transform the rippled layer imbalance according to
\begin{equation}
\delta m_{z1}(x) \rightarrow \delta m_{z1}(\alpha x + \beta y) ,
\end{equation}
then the associated charge density is
\begin{equation}
\delta\rho = -\frac{e\nu_T}{4\pi} Q
              \partial_y \delta m_{z1}(\alpha x + \beta y) ,
\end{equation}
which associates a spatially varying charge density proportional to
$\beta t/U$ to the tilted soliton lines.
The integrated charge of the soliton lines remains zero,
so the shear stiffness\cite{hannasl} may not be greatly affected.

\subsection{Anisotropic transport}

Application of a sufficiently strong in-plane magnetic field produces
soliton lines that are parallel to the in-plane magnetic field.
When the layers are imbalanced, these soliton lines possess
dipole-charge densities.  The resulting electric fields associated
with the dipole-charge densities will make the conductivity anisotropic
in the incommensurate phase.  Transport parallel to the soliton lines
(along the direction of the in-plane magnetic field, say $-{\bf \hat{y}}$)
is expected to be easier than in the direction perpendicular
(along ${\bf \hat{x}}$) to the soliton lines: i.e., when the layers
are imbalanced, we expect
\begin{eqnarray}
\rho_{xx} &\approx& \rho_{yy} , \qquad Q<Q_c ,
\\ \nonumber
\rho_{xx} &>&       \rho_{yy} , \qquad Q>Q_c ,
\end{eqnarray}
Thus a disparity between the longitudinal resistivities parallel and
perpendicular to the in-plane magnetic field indicates the presence of
an incommensurate soliton-lattice phase in the imbalanced system.

We note that anisotropic transport in the quantum Hall regime
has been found in a very different context, in very high-mobility
single-layer samples at high half-integer filling factors.\cite{lilly,du}
In that case, the anisotropic transport is taken to indicate the
presence of a spontaneously striped anisotropic charge density
of the quantum Hall ground state.\cite{stripetheory}
Here, the stripes are not spontaneous (for $t \ll U$),
but are produced and oriented by the in-plane magnetic field.
Nonetheless, the stripes (dipolar soliton lines) found here
should also produce anisotropic transport.

\section{Interlayer capacitance}
\label{sec:capjpe}

Capacitance measurements offer the possibility of directly probing
the ground-state properties of two-dimensional electron systems,
such as the thermodynamic compressibility.
Although the differential gate capacitance, which measures how
the charge on a gate changes with respect to changes in the gate voltage,
is slightly affected by the compressibility of the electron gas
in the occupied layer that is nearest the gate, it is almost entirely
dominated by the large gate-to-layer distance of the device.
A far more sensitive measurement of the electronic compressibility is
provided by the Eisenstein ratio $R_E$ which is an interlayer
capacitance,\cite{jpe,jungwirth,bigdles}
\begin{equation}
\label{eq:redef}
R_E = \delta E_{12} / \delta E_{\rm gate} ,
\end{equation}
where $E_{12}$ is the electric field that exists between the layers,
and $E_{\rm gate}$ is the electric field between the gate and the nearest
layer.  Classically, conduction  electrons in the layers should completely
screen the electric fields produced by the gates, so that $E_{12}$=0
and $R_E$=0. Indeed, this result is approached when the layers are sufficently
far apart (beyond several hundred \AA).  But, due to their finite
density of states, the effectively two-dimensional electron layers
cannot completely screen the gate electric fields,
so $E_{12}$ is nonzero when the gates are unbalanced.

The Eisenstein ratio has been measured experimentally in two-dimensional
electron gas (2DEG) systems, and used to determine the negative
compressibility of the low-density 2DEG;\cite{jpe} it has also been
calculated theoretically at zero magnetic field\cite{jpe,bigdles}
and in 2LQH systems.\cite{jungwirth}
The calculation of the Eisenstein ratio is discussed in some detail in
Refs.~\onlinecite{jpe,jungwirth,bigdles}; here we shall briefly outline
only the key steps.

It is covenient to separate the total energy per unit area of the 2LQH
system into an electrostatic part (i.e., the integrated electrostatic
energy density $\epsilon E_{12}^2d/2$ between the layers) and a many-body part
$\langle\varepsilon\rangle$, which would be the energy per area for a
system with neutralizing charge backgrounds in each layer.\cite{jungwirth}
The chemical potential relative to the bottom of quantum well $j$ is given
by\cite{jpe}
\begin{equation}
\mu_i = \partial \langle\varepsilon\rangle / \partial n_i ,
\end{equation}
where $n_i = \nu_i/2\pi\ell^2$ is the areal number density in layer $i$.
$R_E$ can be expressed in terms of the interlayer separation $d$ and
the effective electronic lengths $s_{ij}$, defined as
\begin{equation}
\label{eq:sij}
s_{ij} = \frac{\epsilon}{e^2}
                     \frac{\partial \mu_i}{\partial n_j} =
\frac{\ell}{2} \frac{\partial^2}{\partial\nu_j\partial\nu_i}
\left( \frac{2\pi\ell^2 \langle\varepsilon\rangle}
            {e^2/4\pi\epsilon\ell} \right) ,
\end{equation}
from which it follows that $s_{ji}=s_{ij}$.
In the absence of interlayer interactions (the case considered in
Ref.~\onlinecite{jpe}), $s_{12}=s_{21}=0$ and the length $s_{ii}$
is inversely proportional to the electronic compressibility $\kappa_i$
in layer $i$ (Ref. \onlinecite{jungwirth}):
\begin{equation}
s_{ii} = \frac{\epsilon}{e^2 n_i^2 \kappa_i} .
\end{equation}

If the front-gate and back-gate voltages are varied simultaneously so
that the total layer density is kept fixed ($\nu_1$+$\nu_2$=1), then
\begin{equation}
\label{eq:res1s2}
R_E \equiv \delta E_{12}/\delta E_{\rm gate} = \frac{s_1+s_2}{d+s_1+s_2} ,
\end{equation}
where
\begin{equation}
s_1 \equiv s_{11} - s_{12} ,
\qquad
s_2 \equiv s_{22} - s_{21} .
\end{equation}
If only one of the gate voltages is kept fixed, then the numerator
of Eq.~(\ref{eq:res1s2}) is equal to either $s_1$ or $s_2$, instead
of their sum.\cite{bigdles}
At high densities, $s_i$ (and thus $R_E$) are positive, but at
suffiently low densities, they become negative, resulting in the
negative values of $R_E$ that have been measured experimentally.\cite{jpe}

The Eisenstein ratio has been calculated for a $\nu_T=1$ 2LQH
state,\cite{jungwirth} although without a parallel magnetic field.
It was found that, although the lengths $s_i$ were negative,
the criterion for stability against abrupt interlayer charge
transfer,\cite{bigdles}
\begin{equation}
\label{eq:ds1s2}
d + s_1 + s_2 > 0 ,
\end{equation}
is still satisfied.
It follows from Eqs. (\ref{eq:res1s2}) and (\ref{eq:ds1s2})
that when $R_E$ diverges, it signals an interlayer charge-transfer
instability.

In the next two subsections, we calculate the contribution $R_{E1}$
of the PT Hamiltonian to the Eisenstein ratio.
$R_{E1}$ contains the dependence of $R_E$ on the tunneling $t$,
pseudospin stiffness $\rho_s$, parallel magnetic field $Q$,
and layer imbalance $m_{z0}$.
It is convenient to separate $R_{E1}$ into two parts:
one that is formally divergent at $Q$=$Q_c$ when the layers are unbalanced
($m_z\ne 0$), and one that does not diverge, but which nevertheless
exhibits a nontrivial dependence on $Q$ and $m_{z0}$.

\subsection{Divergent contribution}
\label{subsec:divergent}

We shall first consider only the contributions to the electronic lengths
$s_i$ that become negative and divergent at the CI transition.
We therefore focus on how changes in the layer imbalance $m_z$ allow
one to cross through the CI transition, due to the $m_z$ dependence of
the critical parallel magnetic field (or of $Q_c$), as discussed in
Sec.~\ref{subsec:parallel} and in Ref.~\onlinecite{hannasl}.
We therefore consider only the PT part of the the energy per area
in Eq.~(\ref{eq:edensity}), and then only that part of the PT energy
that depends on $m_z$ through $Q/Q_c$.
(There are other terms that contribute to $s_i$, but they do not
give rise to a divergence at the CI transition, at least for $t/U \ll 1$,
which is the usual situation in most samples.)
Because we are focusing on the effects of passing throught the CI
transition, rather than the effects of changing the total filling
factor, we restrict our attention to the case of fixed total filling
factor $\nu_T$.
With these restrictions, the effect of taking a derivative of a function
of $Q/Q_c$ with respect to the layer filling factor $\nu_i$
produces the equivalence
\begin{equation}
\label{eq:derivnui}
\frac{\partial}{\partial \nu_i} \rightarrow
-\frac{\partial m_z}{\partial \nu_i} \frac{\partial Q_C}{\partial m_z}
 \frac{Q}{Q_c} \frac{\partial}{\partial Q} =
\frac{(-1)^i}{2} \frac{m_z}{1-m_z^2} Q \frac{\partial}{\partial Q} .
\end{equation}
Combining Eqs. (\ref{eq:sij}) and (\ref{eq:derivnui}) with the magnetization
calculations of Ref.~\onlinecite{hannasl} gives
\begin{equation}
\label{eq:s1pluss2}
\frac{s_1+s_2}{\ell} \sim
-\pi \left( \frac{m_z}{1-m_z^2} \right)^2 (Q\ell)^2
\frac{\rho_s}{e^2/4\pi\epsilon\ell}
\frac{\partial Q_s}{\partial Q} ,
\end{equation}
where we have kept only the contribution that diverges at the CI transition.
The divergence occurs because nonzero $m_z$ allows small changes in
the layer filling factors to tune the system through the CI transition.
The divergent part of the electronic lengths $s_i$ are thus proportional
to the in-plane differential magnetic susceptibility that was calculated in
Ref.~\onlinecite{hannasl}, and which diverges at the CI transition.

Unfortunately, the divergence in the electronic lengths $s_i$ at the
CI transition is weak, proportional to $1/\sqrt{Q-Q_c}$, with a small
prefactor.  For the hypothetical sample considered in this paper and
$m_{z0}$=0.5, Eqs. (\ref{eq:newqs}) and (\ref{eq:s1pluss2}) give
\begin{equation}
\frac{s_1+s_2}{\ell} \sim -\frac{0.005}{\sqrt{Q/Q_c-1}} ,
\end{equation}
which requires $(Q/Q_c-1) \sim 10^{-5}$ to give a divergent value of
the Eisenstein ratio (i.e., $d+s_1+s_2=0$), at least for a
tunneling-matrix element $t_0$=0.5~meV.  The required nearness to the CI
transition (i.e., the smallness of $Q/Q_c-1$) is inversely proportional
to $t_0$, so making $t_0$ smaller might be of some help.
However, it may be that in practice, sample disorder smears out the
CI transition well before the divergence in $R_E$ can be approached.

\subsection{Nondivergent contributions}

We saw in Sec.~\ref{subsec:divergent} that, although there is a contribution
to the interlayer capacitance (the Eisenstein ratio $R_E$) which is
formally divergent at the CI transition, it may be difficult in
practice to tune close enough to the transition to observe the
divergence.  It may be that the nondivergent contribution of the
PT contribution to $R_E$ is more easily detected.  We calculate
this contribution below for small $2t_0/U$.

For simplicity, we focus on the case of fixed total filling factor.
Then it follows from the definition of $R_E$ in Eq.~(\ref{eq:redef}),
from Gauss' law, from the definition of $V_g$ in Eq.~(\ref{eq:vg}),
and from Eq.~(\ref{eq:mz}), that we may express $R_E$ as
\begin{eqnarray}
R_E &=& 1 - \frac{\bar{D}_1}{U}
\frac{\partial \langle m_z \rangle}{\partial m_{z0}} \approx
R_{E0} + R_{E1} ,
\\ \nonumber
R_{E0} &=& 1 - \bar{D}_1/U ,
\\ \nonumber
R_{E1} &=&  -\frac{\bar{D}_1}{U}
            \frac{\partial \langle m_{z1} \rangle}{\partial m_{z0}}
\\ \nonumber &=& \frac{2t_0}{U}
\left[
  \frac{\langle \cos\tilde{\theta}_0 \rangle}{(1-m_{z0}^2)^{-3/2}}
- \xi_0^2 \langle (\nabla \tilde{\theta}_0 - {\bf Q})^2 \rangle
\right] .
\end{eqnarray}
The largest contribution to the Eisenstein ratio is $R_{E0}$, and it
was this quantity that was calculated in Ref.~\onlinecite{jungwirth}.
For the hypothetical sample parameters in the text, $R_{E0} = -0.9$,
independent of $m_{z0}$ and $Q$.
$R_E=R_{E0}$ when the interlayer-tunneling amplitude $t_0$ is very
small, or when the parallel field is large ($Q$ substantially larger
than $Q_c$).

$R_{E1}$ contains all the dependence of $R_E$ on the in-plane
magnetic field and layer imbalance.
If $t/U$ is made as large as possible, then by measuring the
$Q$ and $m_{z0}$ dependence of $R_E$, it may be possible to
measure $R_{E1}$.
In the commensurate phase ($Q<Q_c$), $\tilde{\theta}_0=0$, so
\begin{equation}
R_{E1} = -\frac{\bar{D}_1}{U} \frac{2t_0/U}{\sqrt{1-m_{z0}^2}}
\left[ \frac{1}{(1-m_{z0}^2)} -
       \left( \frac{4}{\pi} \frac{Q}{Q_c} \right)^2 \right] .
\end{equation}
In the absence of an in-plane magnetic field ($Q$=0),
\begin{equation}
R_{E1}(Q=0) = -\frac{\bar{D}_1}{U}
               \frac{2t_0/U}{\sqrt{1-m_{z0}^{3/2}}}
\approx -0.39 ,
\end{equation}
for the hypothetical sample parameters described in the text,
with $m_{z0}$=0.5, so that in this case
$R_{E1}$ is almost half the size of $R_{E0}$.
As $Q$ increases, $R_{E1}$ decreases in magnitude and,
provided that $m_{z0} < \sqrt{1-(\pi/4)^2} \approx 0.62$,
passes through zero and becomes positive near the CI transition.

At the CI transition, this nondivergent part of $R_{E1}$ is
\begin{equation}
R_{E1}(Q=Q_c) = \frac{\bar{D}_1}{U} \frac{2t_0/U}{\sqrt{1-m_{z0}^2}}
\left[ \left( \frac{4}{\pi} \right)^2 -
\frac{1}{(1-m_{z0}^2)} \right] ,
\end{equation}
which for the sample parameters described in the text has a value of
$R_{E1}(Q=Q_c)$=0.16 for $m_{z0}$=0 and 
$R_{E1}(Q=Q_c)$=0.08 for $m_{z0}$=0.5.
In the incommensurate phase, $R_{E1}$ drops rapidly to zero due to
rapid spatial variations in $\tilde{\theta}_0(x)$.\cite{hannasl}
By measuring how the in-plane magnetic field and layer imbalance
affect the interlayer capacitance $R_E$, it may be possible to
estimate the pseudospin stiffness $\rho_s$ in the commensurate phase,
and also to detect the incommensurate phase, as signaled by a
rapid decrease in the sensitivity of $R_E$ to $Q$ and $m_{z0}$.

\section{Conclusions}

It has been shown that the incommensurate phase of a bilayer quantum Hall
state has a ``rippled'' dipole charge density whenever the layers are
unbalanced.
The rippling arises because the layer imbalance $m_z$ and the interlayer
phase $\theta$ are coupled through the $m_z$ dependence of the
effective tunneling energy $t$ and pseudospin stiffness $\rho_s$
in the Pokrovsky-Talapov part of the total energy in Eq.~(\ref{eq:edensity}).
This coupling between the layer imbalance and interlayer phase
produces the rippled state when the translational symmetry of the
phase is broken in the incommensurate phase.
The rippled layer imbalance was calculated within the
Hartree-Fock gradient approximation, and is illustrated in
Fig.~\ref{fig:ripple}.  The details of the calculation are given
in the Appendix.
We focused on the limit where the interlayer-tunneling energy
is smaller than the charging energy of the bilayer ($t\ll U$),
which is the case for all bilayer samples which have been studied
experimentally so far.

Because solitons have an associated dipole-moment per unit length
(which we estimated in the Appendix) in the rippled state,
well-separated soliton lines experience a power-law (inverse cube)
repulsive-force per unit length, rather than the much weaker
exponentially-decaying repulsion between solitons found in the
balanced case.
This has a strong effect on how the density of solitons depends on
the in-plane magnetic field near the CI transition,
as illustrated in Fig.~\ref{fig:qs}.

The fact that solitons have an electric dipole moment
in the rippled state makes it likely that they produce anisotropic
transport.  Transport parallel to the soliton lines ($\pm{\bf \hat{y}}$)
is likely to be easier than transport perpendicular to the soliton lines
($\pm{\bf \hat{x}}$).  We expect the ratio $\rho_{xx}/\rho_{yy}$ to increase
when the rippled state is entered.  This could provide an experimental
signature for the incommensurate state.

We calculated the interlayer capacitance, specifically the
Eisenstein ratio $R_E$, which is a sensitive measure of the
electronic compressibility and of interlayer electronic correlations.
When the layers are unbalanced, there is a contribution to $R_E$
which diverges at the CI transition.
However, observing this contribution is likely to be problematic because
it requires that the CI transition be very sharp (unsmeared by disorder).
We also calculated a contribution to $R_E$ which does not diverge at
the CI transition, but which nevertheless offers a good possibility of
experimental detection.  We calculated the in-plane magnetic-field dependence
of $R_E$, along with its dependence on layer imbalance, especially
in the commensurate ($Q<Q_c$) phase, up to the CI transition ($Q$=$Q_c$),
and discussed its behavior (rapid decline) in the incommensurate
($Q>Q_c$) phase.
By measuring $R_E$, the pseudospin stiffness could be estimated and the
incommensurate phase detected.

The existence of ``rippled'' layer densities in
the incommensurate phase of unbalanced 2LQH systems
is expected to be valid beyond the HFGA,
although the size of the density variations will be reduced
by quantum fluctuations.\cite{moonfluct,yogfluct}
As long as the layer densities are ``rippled'',
they will make an anomalous contribution to the capacitance.
Although fluctuation effects change the sizes of $\rho_s$ and
$t$,\cite{moonfluct,yogfluct} the basic physics
of producing ``rippled'' layer densities still holds beyond the HFA.
In particular, $\rho_s$ and $t$ will still change with the layer
imbalance $m_z$, and this dependence on $m_z$ will produce a coupling
between $m_z$ and the pseudospin angle $\theta$,
leading to nonuniform $m_z$ in the incommensurate phase when $\nu_1\ne\nu_2$.

Several remarks about the observability of the capacitive effects
found here are in order, especially the effects of including
finite temperature and disorder.
In practice, both finite temperature and disorder limit the minimum
effective value of $(Q/Q_c-1)$ that can be obtained.
These are important topics which deserve further study;
only some preliminary considerations are discussed here.

Ignoring the effects of disorder for the moment, it is important to
note that the soliton lattice exists only at sufficiently
low temperatures.
The soliton lattice supports a finite-temperature Kosterlitz-Thouless
(KT) transition due to dislocation-mediated melting of the lattice
of soliton lines.\cite{dennijs,papapark}
The KT temperature for melting the soliton lattice is roughly
$k_BT_{KT} \sim (\pi/2) \sqrt{K_1K_2}$,
where $K_1$ is the longitudinal stiffness and $K_2$ is the transverse
stiffness of the soliton lattice.\cite{read}
Because $K_1 \rightarrow 0$ as $Q \rightarrow Q_c$, the KT temperature
drops as the CI transition is approached.
This sets a limit to how close one can get to the CI transition,
which may be estimated in the case of finite layer imbalance by using
$K_1=2\sqrt{C(Q/Q_c-1)}$ [see Eq.~(\ref{eq:newqs})]
and $K_2=\rho_s$.\cite{hannasl}
From these considerations, the requirement that $T<T_{KT}$ gives
\begin{equation}
(Q/Q_c-1) > \frac{1}{4} \left( \frac{2}{\pi} \frac{k_BT}{\rho_s} \right)^4
          \approx 2.4 \left( \frac{T}{1~{\rm K}} \right)^4
\end{equation}
for the sample parameters used in the text.
For $T$=100~mK, this yields $(Q/Q_c-1) > 2.4\times 10^{-4}$.

The smearing of the CI transition due to disorder is more problematic.
Even in capacitance experiments designed to measure the
(in)compressibility of the fractional quantum Hall (FQH) state
(which would, in the absence of disorder, lead to a divergent $R_E$
at odd-denominator filling factors), only finite changes in $R_E$ are
found at the FQH filling factors, due to the effects of disorder.\cite{jpe}
It is to be expected that disorder may eliminate any abrupt features
in the interlayer capacitance $R_E$ in this case also.
Strictly speaking, the long-range order of the SL is destroyed by
any finite amount of disorder;\cite{mpaf}
presumably the SL has only a finite correlation length
$\zeta_d<\infty$ due to disorder.
Roughly speaking, the maximum spacing between solitons ($L_s$) will
be limited to $L_s<\zeta_d$, so that
\begin{equation}
(Q/Q_c-1) > C \left( \frac{\pi^2}{2} \frac{\xi}{\zeta_d} \right)^2 .
\end{equation}
In practice, such disorder could arise from small variations in the
local tunneling ampitudes $t$ or the spin stiffness $\rho_s$
due to minute variations in the interlayer barrier thickness and/or
the layer separation.
More theoretical work needs to be done to determine the limits
imposed by disorder, but the issue of the observability of the
CI transition in capacitance measurements must be settled experimentally.

\section*{Acknowledgments}

The author gratefully acknowledges many illuminating discussions with
A.~H.~MacDonald and S.~M.~Girvin.
It is a pleasure to thank A.~R.~Hamilton for answering several questions
regarding experimental measurements on double-layer systems.
Thanks are also due to R.~J.~Reimann and G.~R.~Newby
for useful discussions.
This work was supported by an award from Research Corporation
and by NSF Grant No. DMR-997232.  The author is thankful for
the support provided by the ITP Scholars Program at U. C. Santa
Barbara, where part of this work was carried out: thus this
research was also supported in part by the National Science Foundation
under Grant No. PHY99-07949.
In addition, the work described is supported by the NSF-Idaho EPSCoR Program
and by the National Science Foundation under Grant No. EPS-0132626.

\appendix
\section*{Dipole moment}

In this appendix, we estimate the dipole-moment per unit length associated
with solitons in the incommensurate phase of an imbalanced 2LQH system.
Expanding Eq.~(\ref{eq:edensity}) in powers of $(t/U)$ gives
\begin{equation}
m_z({\bf r}) \approx m_{z0} + m_{z1}({\bf r}) ,
\end{equation}
where
\begin{equation}
m_{z0} = V_g/U
\end{equation}
is the $t$=0 result for the layer imblance, and
\begin{equation}
m_{z1}({\bf r}) = \frac{2t}{U} \frac{m_{z0}}{(1-m_{z0}^2)}
\left[ \xi^2 (\nabla \tilde{\theta}_0 - {\bf Q})^2
       - \cos\tilde{\theta}_0 \right] ,
\end{equation}
to first order in $t/U$.
Here $\tilde{\theta}_0({\bf r})$ is the soliton-line solution to
the PT model for the values of $t$ and $\rho_{\rm s}$ corresponding to
setting $m_z$=$m_{z0}$ in Eq.~(\ref{eq:deftrhos}).

\begin{figure}[h]
\epsfxsize3.5in
\centerline{\hspace{0.3in}\epsffile{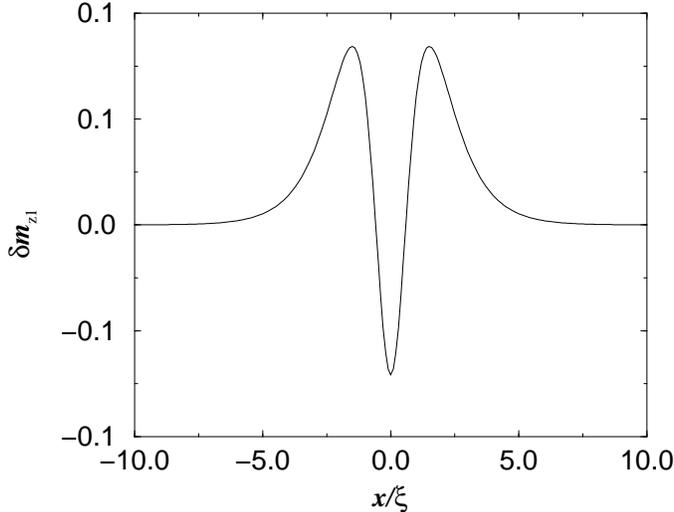}}
\caption{
Spatial dependence of the layer imbalance $\delta m_{z1}(x)$ associated
with a single soliton, vs position for the hypothetical ``typical''
sample parameters described in the text.
}
\label{fig:onedeltamz}
\end{figure}

For a single soliton, the lowest-order solution for $\tilde{\theta}$
is\cite{hannasl}
\begin{equation}
\tilde{\theta}_0(x) = 4 \arctan \exp(x/\xi) ,
\end{equation}
where we have taken ${\bf Q}$ to lie along the $\hat{\bf x}$ direction.
This gives
\begin{eqnarray}
1 - \cos\tilde{\theta}_0(x) &=& \frac{2}{\cosh^2(x/\xi)} ,
\\ \nonumber
\xi\partial_x \tilde{\theta}_0(x) &=& \frac{2}{\cosh(x/\xi)} .
\end{eqnarray}
It is convenient to express $m_{z1}({\bf r})$ as
\begin{equation}
m_{z1}({\bf r}) = \bar{m}_{z1} + \delta m_{z1}(x) ,
\end{equation}
where
\begin{equation}
\bar{m}_{z1} = \frac{2t}{U} \frac{m_{z0}}{(1-m_{z0}^2)} (Q^2\xi^2 - 1)
\end{equation}
is the value of $m_{z1}$ in the commensurate ($\tilde{\theta}_0 = 0$) phase.
For the hypothetical sample described in the text,
$\bar{m}_{z1} \approx 0.048$ for $m_{z0}$=0.5 and $Q \approx Q_c$.
We associate the spatially dependent part of $m_z$
(see Fig.i~\ref{fig:onedeltamz}) with the soliton line:
\begin{equation}
\delta m_{z1}(x) = \frac{2t}{U} \frac{m_{z0}}{(1-m_{z0}^2)}
\left[ \frac{4Q\xi}{\cosh(x/\xi)}
       - \frac{6}{\cosh^2(x/\xi)} \right] .
\end{equation}

The areal number density of layer $j$ is given by
$n_j = \nu_j / (2\pi\ell^2)$.
Therefore, the dipole-moment per unit area is
\begin{equation}
\frac{p}{L_x L_y} = -\frac{ed}{2\pi\ell^2} m_z ,
\end{equation}
and we associate an dipole-moment per unit length
\begin{eqnarray}
\frac{\delta p}{\delta y} &=& -\frac{ed}{2\pi\ell^2}
\int_{-\infty}^\infty \delta m_{z1}(x) dx
\\ \nonumber &=&
-\frac{ed\xi}{2\pi\ell^2} \frac{8t}{U} \frac{m_{z0}}{(1-m_{z0}^2)}
(\pi Q\xi - 3)
\\ \nonumber &=&
-\frac{ed\xi_0}{2\pi\ell_0^2} \frac{2t_0}{U}
\frac{m_{z0}}{(1-m_{z0}^2)^{1/4}}
(4Q/Q_c - 3) .
\end{eqnarray}
For the ``typical''sample described in Sec.~\ref{subsec:effham} at
layer imbalance $m_{z0}=1/2$ and $Q$=$Q_c$,
\begin{equation}
\delta p/\delta y \approx -0.10 e ,
\end{equation}
where $-e$ is the electric charge of an electron.

When the soliton lines do not overlap ($Q$ sufficiently near $Q_c$),
then $\tilde{\theta}_0(x)$ is very nearly a periodic superposition of
single-soliton solutions, spaced apart by $L_{\rm s}$:
\begin{equation}
\tilde{\theta}_0({\bf r}) \approx
4 \sum_j \arctan \exp [(x-j L_{\rm s})/\xi] .
\end{equation}
The dipolar interaction-energy per unit length between two
parallel soliton lines separated by a distance $x$ is
\begin{equation}
\frac{{\cal V}_2(x)}{L_y} = \frac{(\delta p/\delta y)^2}{2\pi\epsilon x^2} .
\end{equation}
Thus the total dipole-interaction energy per unit area is
\begin{eqnarray}
\frac{{\cal V}}{L_xL_y} &=&
\frac{N_{\rm s}}{L_xL_y} \sum_{j=1}^{\infty} {\cal V}_2(jL_{\rm s})
\\ \nonumber &=&
\frac{\pi}{12} \frac{(\delta p/\delta y)^2}{\epsilon L_{\rm s}^3} =
\frac{1}{96\pi^2} \frac{(\delta p/\delta y)^2}{\epsilon} Q_{\rm s}^3 ,
\end{eqnarray}
where we have used the fact that the number of solitons is
$N_{\rm s} = L_z/L_{\rm s}$.

The relation between the wave vector $Q$ and the parameter $\eta$
is obtained by minimizing the total energy per unit area with respect
to $Q_s$ at fixed $Q$;\cite{hannasl} when the layers are balanced,
Eq.~(\ref{eq:qqc}) results, and Eqs. (\ref{eq:qsqc}) and (\ref{eq:qqc})
may be combined to obtain $Q_s$ as a function of $Q$.
When the layers are imbalanced,
Eq.~(\ref{eq:qqc}) acquires an additional term due to
the dipole interactions between solitons,
\begin{eqnarray}
\label{eq:qqcdipole}
Q/Q_c &=& E(\eta)/\eta +
\frac{1}{\rho_s Q_c} \frac{\partial}{\partial Q_s}
\frac{{\cal V}}{L_xL_y}
\\ \nonumber &=&
E(\eta)/\eta + C (Q_s/Q_c)^2 ,
\end{eqnarray}
where
\begin{equation}
C = \frac{1}{8\pi} \left( \frac{\delta p/\delta y}{e} \right)^2
\frac{e^2/4\pi\epsilon\ell}{\rho_s} Q_c\ell ,
\end{equation}
and $C \sim 0.14$ for the hypothetical ``typical'' sample with
$m_{z0}$=0.5 and $Q$=$Q_c$.
Because the interaction between separated solitons is an inverse-power law
(when unbalanced) rather than exponetially decaying function (when balanced),
$Q_s$ and $K_1$ are proportional to $\sqrt{Q-Q_c}$ near the CI transition.


\begin{thebibliography}{99}

\bibitem{tsui} D.C. Tsui, H.L. St\"ormer, and A.C. Gossard,
Phys. Rev. Lett. {\bf 48}, 1559 (1982).

\bibitem{prange} {\it The Quantum Hall Effect},
edited by R.E. Prange and S.M. Girvin
(Springer-Verlag, New York, 1990),
and references therein.

\bibitem{laughlin} R.B. Laughlin,
Phys. Rev. Lett. {\bf 50}, 1395 (1983).

\bibitem{sondhi} S.L. Sondhi, A. Karlhede, S.A. Kivelson,
and E. H. Rezayi,
Phys. Rev. B {\bf 47}, 16 419 (1993).

\bibitem{gmchap} See S.M. Girvin and A.H. MacDonald,
in {\it Perspectives in Quantum Hall Effects},
edited by S. Das Sarma and A. Pinczuk (Wiley, New York, 1997),
and references therein.

\bibitem{gruner} G. Gr\"uner,
{\it Density Waves in Solids} (Addison-Wesley, New York, 1995),
and references therein.

\bibitem{bak} P. Bak,
Rep. Prog. Phys. {\bf 45}, 587 (1982).

\bibitem{dennijs} Marcel den Nijs, in
{\it Phase Transitions in Critical Phenomena},
edited by C. Domb and J. L. Lebowitz (Academic Press, New York, 1988),
Vol. 12, pp. 219-333.

\bibitem{yang} K. Yang, K. Moon, L. Zheng, A.H. MacDonald,
S.M. Girvin, D. Yoshioka, and S.C. Zhang,
Phys. Rev. Lett. {\bf 72}, 732 (1994).

\bibitem{moon} K. Moon, H. Mori, K. Yang, S.M. Girvin,
A.H. MacDonald, L. Zheng, D. Yoshioka, and S.C. Zhang,
Phys. Rev. B {\bf 51}, 5138 (1995).

\bibitem{order} S.M. Girvin and A.H. MacDonald,
Phys. Rev. Lett. {\bf 58}, 1252 (1987).

\bibitem{murphy} S.Q. Murphy, J.P. Eisenstein, G.S. Boebinger,
L.N. Pfeiffer, and K.W. West,
Phys. Rev. Lett. {\bf 72}, 728 (1994).

\bibitem{read} N. Read,
Phys. Rev. B {\bf 52}, 1926 (1995).

\bibitem{mullen} K. Moon and K. Mullen,
Phys. Rev. B {\bf 57}, 1378 (1998).

\bibitem{ep2ds12} C.B. Hanna, A.H. MacDonald, and S.M. Girvin,
Physica B {\bf 249-251}, 824 (1998).

\bibitem{hannasl} C.B. Hanna, A.H. MacDonald, and S.M. Girvin,
Phys. Rev. B {\bf 63}, 125305 (2001).

\bibitem{hannamar97} C.B. Hanna,
Bull. Am. Phys. Soc. {\bf 42}, 553 (1997).

\bibitem{jogunequal}  Y.N. Joglekar and A.H. MacDonald,
cond-mat/0111056 (unpublished).

\bibitem{radzcant} Leo Radzihovsky,
Phys. Rev. Lett. {\bf 87}, 236802 (2001).

\bibitem{abolglobal} M. Abolfath, L. Radzihovsky, and A.H. MacDonald,
cond-mat/0110049 (unpublished).

\bibitem{dlexpt} J.P. Eisenstein, G.S. Boebinger, L.N. Pfeiffer,
K.W. West, and S. He,
Phys. Rev. Lett. {\bf 68}, 1383 (1992);
Y.W. Suen, L.W. Engel, M.B. Santos,
M. Shayegan, and D.C. Tsui,
{\it ibid.} {\bf 68}, 1379 (1992).

\bibitem{ahmpaper} A.H. MacDonald,
Surf. Sci. {\bf 229}, 1 (1990).

\bibitem{bigyang} K. Yang, K. Moon, L. Belkhir, H. Mori,
S.M. Girvin, A.H. MacDonald, L. Zheng, and D. Yoshioka,
Phys. Rev. B {\bf 54}, 11 644 (1996).

\bibitem{spielman} I.B. Spielman, J.P. Eisenstein, L.N. Pfeiffer,
and K.W. West,
Phys. Rev. Lett. {\bf 84}, 5808 (2000);
{\bf 87}, 036803 (2001). 

\bibitem{pt} V.L. Pokrovsky and A.L. Talapov,
Phys. Rev. Lett. {\bf 42} 65 (1970);
Zh. Eksp. Teor. Fiz. {\bf 78} 269 (1980)
[Sov. Phys. JETP {\bf 51} 134 (1980)].

\bibitem{hu} J. Hu and A.H. MacDonald,
Phys. Rev. B {\bf 46}, 12 554 (1992).

\bibitem{moonfluct} K. Moon,
Phys. Rev. Lett. 78, 3741 (1997). 

\bibitem{yogfluct} Yogesh N. Joglekar and Allan H. MacDonald,
Phys. Rev. B {\bf 64}, 155315 (2001).

\bibitem{hamilton} A.R. Hamilton, M.Y. Simmons, F.M. Bolton,
N.K. Patel, I.S. Millard, J.T. Nicholls, D.A. Ritchie, and M. Pepper,
Phys. Rev. B {\bf 54}, R5259 (1996).

\bibitem{sawada} A. Sawada, Z.F. Ezawa, H. Ohno, Y. Horikoshi, A. Urayama,
Y. Ohno, S. Kishimoto, F. Matsukura, and N. Kumada,
Phys. Rev. B {\bf 59}, 14888 (1999);
A. Sawada, Z.F. Ezawa, H. Ohno, Y. Horikoshi, O. Sugie,
S. Kishimoto, F. Matsukura, Y. Ohno, and M. Yasumoto,
Solid State Commun. {\bf 103}, 447 (1997).

\bibitem{gr} I.S. Gradshteyn and I.M. Ryzhik,
{\it Table of Integrals, Series, and Products}
(Academic Press, New York, 1980), Sec. 8.1.

\bibitem{lilly} M.P. Lilly, K.B. Cooper, J.P. Eisenstein,
L.N. Pfeiffer, and K.W. West,
Phys. Rev. Lett. {\bf 82}, 394 (1999);
{\bf 83}, 824 (1999).

\bibitem{du} R.R. Du, D.C. Tsui, H.L. St\"{o}rmer, L.N. Pfeiffer,
K. W. Baldwin, and K. W. West,
Solid State Commun. {\bf 109}, 389 (1999).

\bibitem{stripetheory} A.A. Koulakov, M.M. Fogler, and B.I. Shklovskii,
Phys. Rev. Lett. {\bf 76}, 499 (1996);
M.M. Fogler, A.A. Koulakov, and B.I. Shklovskii,
Phys. Rev. B {\bf 54}, 1853 (1996).

\bibitem{jpe} J.P. Eisenstein, L.N. Pfeiffer, and K.W. West,
Phys. Rev. B {\bf 50}, 1760 (1994);
Phys. Rev. Lett. {\bf 68}, 674 (1992).

\bibitem{jungwirth} T. Jungwirth and A.H. MacDonald,
Phys. Rev. B {\bf 53}, 9943 (1996).

\bibitem{bigdles} C.B. Hanna, Dylan Haas, and J.C. D\'{\i}az-V\'{e}lez,
Phys. Rev. B {\bf 61}, 13882 (2000).

\bibitem{papapark} Thermal fluctuations will renormalize the SL
stiffness and lead to other interesting effects; for recent work, see:
E. Papa and A. Tsvelik, cond-mat/0201343 (unpublished);
S. Park, K. Moon, C. Ahn, J. Yeo, C. Rim, and B.H. Lee,
cond-mat/0203498 (unpublished).

\bibitem{mpaf} M.P.A. Fisher (private communication).

\end{thebibliography}
\end{document}